\pdfoutput=1
\documentclass[onecollarge]{svjour2}
\usepackage{bm}
\usepackage{amsmath}
\usepackage{amssymb}
\usepackage{graphicx}
\usepackage{epsfig}
\usepackage{color}

\newcommand\nn{\nonumber}
\newcommand\bea{\begin{eqnarray}}
\newcommand\eea{\end{eqnarray}}
\newcommand\f{\frac}
\newcommand\p{\partial}
\newcommand\la{\langle}
\newcommand\ra{\rangle}

\begin{document}

\title{Tagged Particle Diffusion in One-Dimensional Systems with Hamiltonian Dynamics -- II}

\author{Anjan Roy \and Abhishek Dhar \and Onuttom Narayan \and Sanjib Sabhapandit}

\institute{A. Roy \at Raman Research Institute, Bangalore 560080, India; anjanroy@rri.res.in
\and A. Dhar \at International Centre for Theoretical Sciences, TIFR, Bangalore 560012, India
\and O. Narayan \at Department of Physics, University of California, Santa Cruz, California 95064, USA
\and S. Sabhapandit \at Raman Research Institute, Bangalore 560080, India
}

\date{\today}

\maketitle

\begin{abstract}
We study various temporal correlation functions of a tagged particle
in one-dimensional systems of interacting point particles evolving  with
Hamiltonian dynamics. Initial conditions of the particles are chosen from the canonical thermal distribution. 
The correlation functions are studied in finite systems, and their forms examined at short and long times.
Various one-dimensional systems are studied. Results of numerical simulations 
for the Fermi-Pasta-Ulam chain are qualitatively similar to results for  the harmonic chain, 
and agree unexpectedly well with a simple description in terms of linearized equations for damped fluctuating sound waves. 
Simulation results for the alternate mass hard particle gas reveal that --- 
in contradiction to our earlier results~\cite{roy13} with smaller system sizes --- the diffusion 
constant slowly converges to a constant value, in a manner consistent with mode coupling theories. 
Our simulations also show that the behaviour of the Lennard-Jones gas depends on its density. 
At low densities, it behaves like a hard-particle gas, and at high densities like an anharmonic chain.  In all the systems studied, the 
tagged particle was found to show normal diffusion asymptotically, with convergence times depending on the system under study. 
Finite size effects show up at time scales larger than sound traversal times, their nature being system-specific. 

\keywords{Hamiltonian dynamics \and 1-d system \and tagged particle
diffusion \and velocity auto-correlation function (VAF) \and mean-squared displacement (MSD)}
\end{abstract}

\section{Introduction}
\label{sec:intro}

\begin{figure}
\includegraphics[scale=0.55]{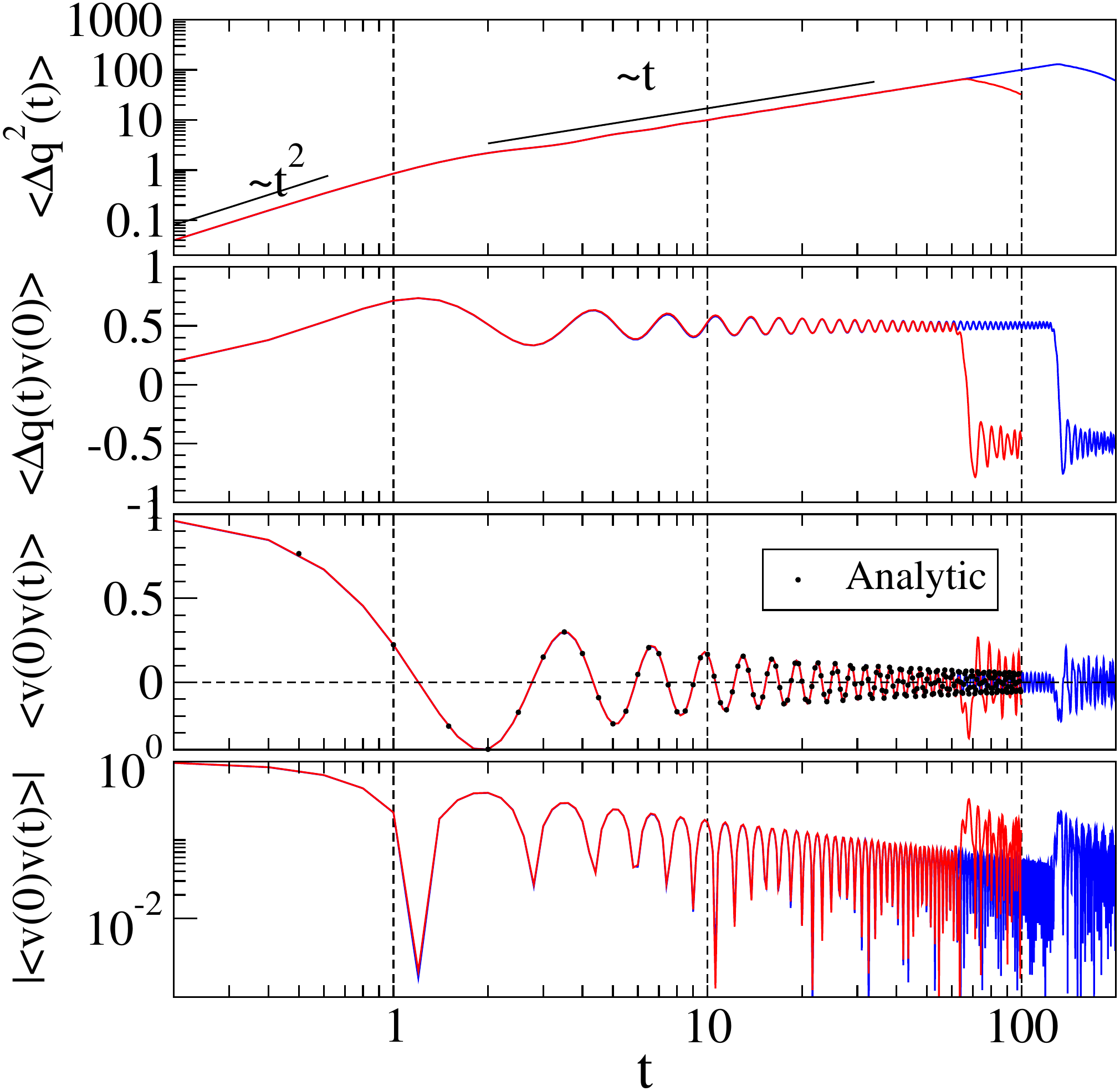}
\caption{ Harmonic chain: short time correlation functions of the central
tagged particle in a system of size $N=65$ (red) and $129$ (blue). Fixed boundary conditions were used and 
the parameters were taken as $L=N, k=1$, $m=1$ and $k_B T=1$.  The diffusion constant can be seen to saturate to the expected value $k_BT/(2 \rho c)=0.5$.  For the VAF, we have also  
plotted the analytic result $\la v(0) v(t) \ra = J_0(2 t)$ }
\label{harmdat}
\end{figure}

\begin{figure}
\includegraphics[scale=0.5]{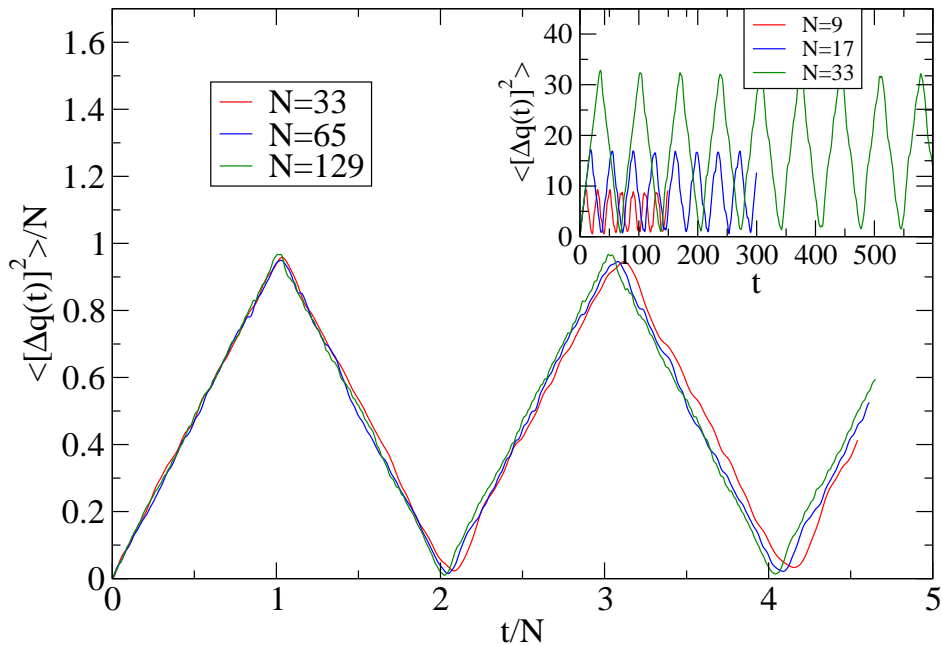}
\caption{Harmonic chain: long time  MSD of the central tagged particle in systems 
of different sizes, and  with fixed boundary conditions. The parameters here are same as in Fig.~(\ref{harmdat}). 
Note the near recurrent behaviour of the MSD.}
\label{harm}
\end{figure}

\begin{figure}
\includegraphics[scale=0.55]{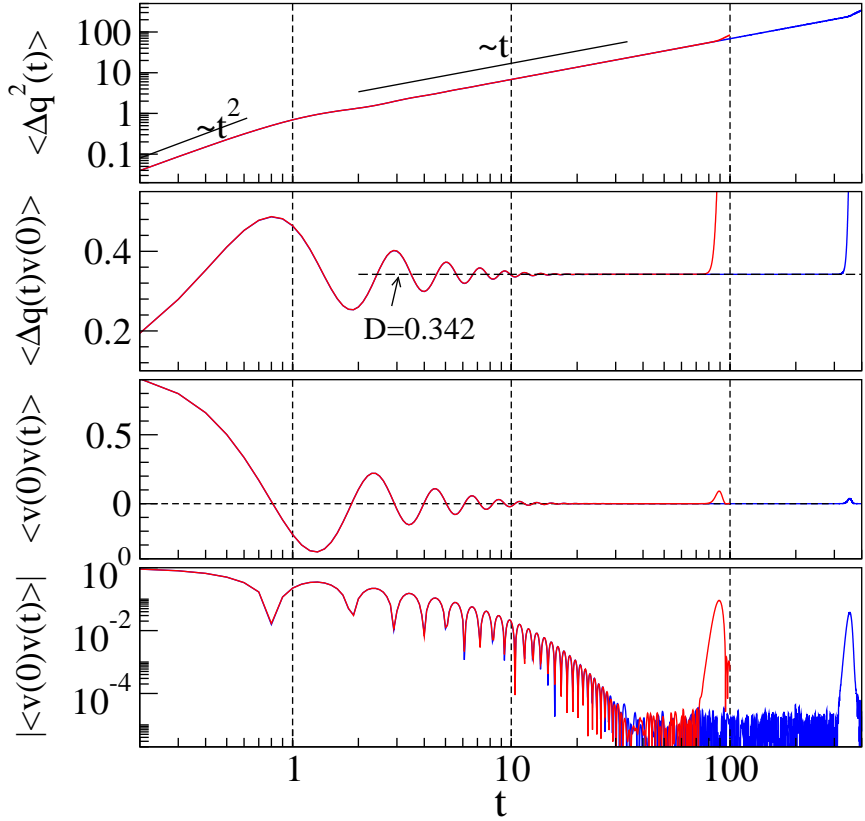}
\caption{FPU chain: short time correlation functions of the central tagged particle
in systems of sizes $N=129$ (red) and $513$ (blue) with periodic boundary conditions. 
The parameters here were taken as $L=N, k=1, \nu =1$, $m=1$ and $k_B T=1$. We see that there is 
a fast convergence of $\la \Delta q(t) v(0) \ra$ to the expected diffusion constant $D=0.342$.
Note that, because of the use of periodic boundaries, the curves go up (positive correlation) when the wall effect sets in, rather than going down [anti-correlation, see Fig. (\ref{harmdat})].}
\label{fpudat}
\end{figure} 

\begin{figure}
\includegraphics[scale=0.5]{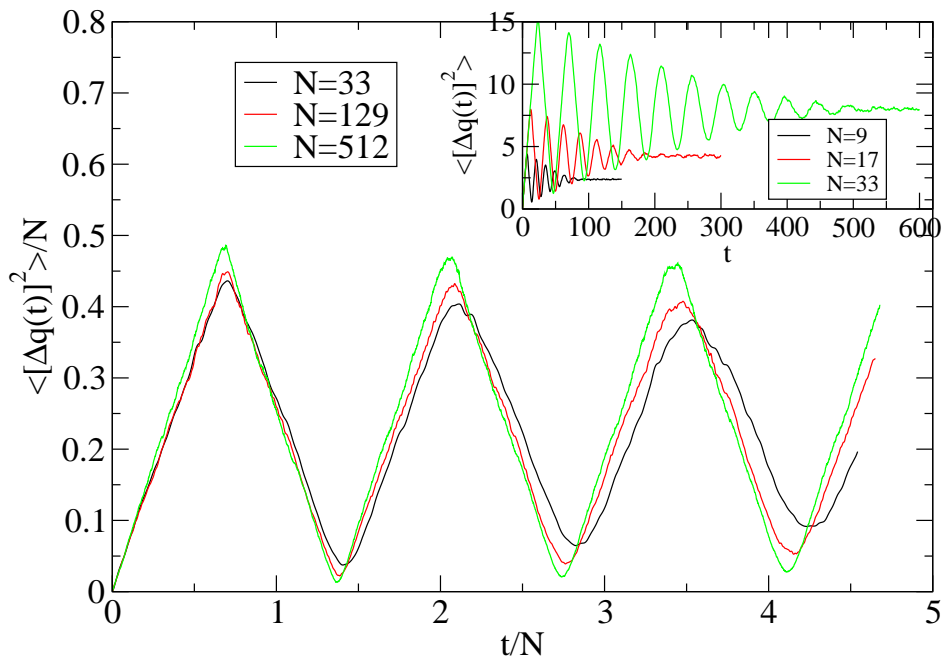}
\caption{ FPU chain: long time MSD of the central tagged particle in FPU chains of different sizes with fixed boundary conditions. The parameters here are the same as in Fig.~(\ref{fpudat}). The sound speed in this case, as calculated within the effective model (see main text), is $c\approx 1.46$.}
\label{fpudat2}
\end{figure}

\begin{figure}
\includegraphics[scale=0.55]{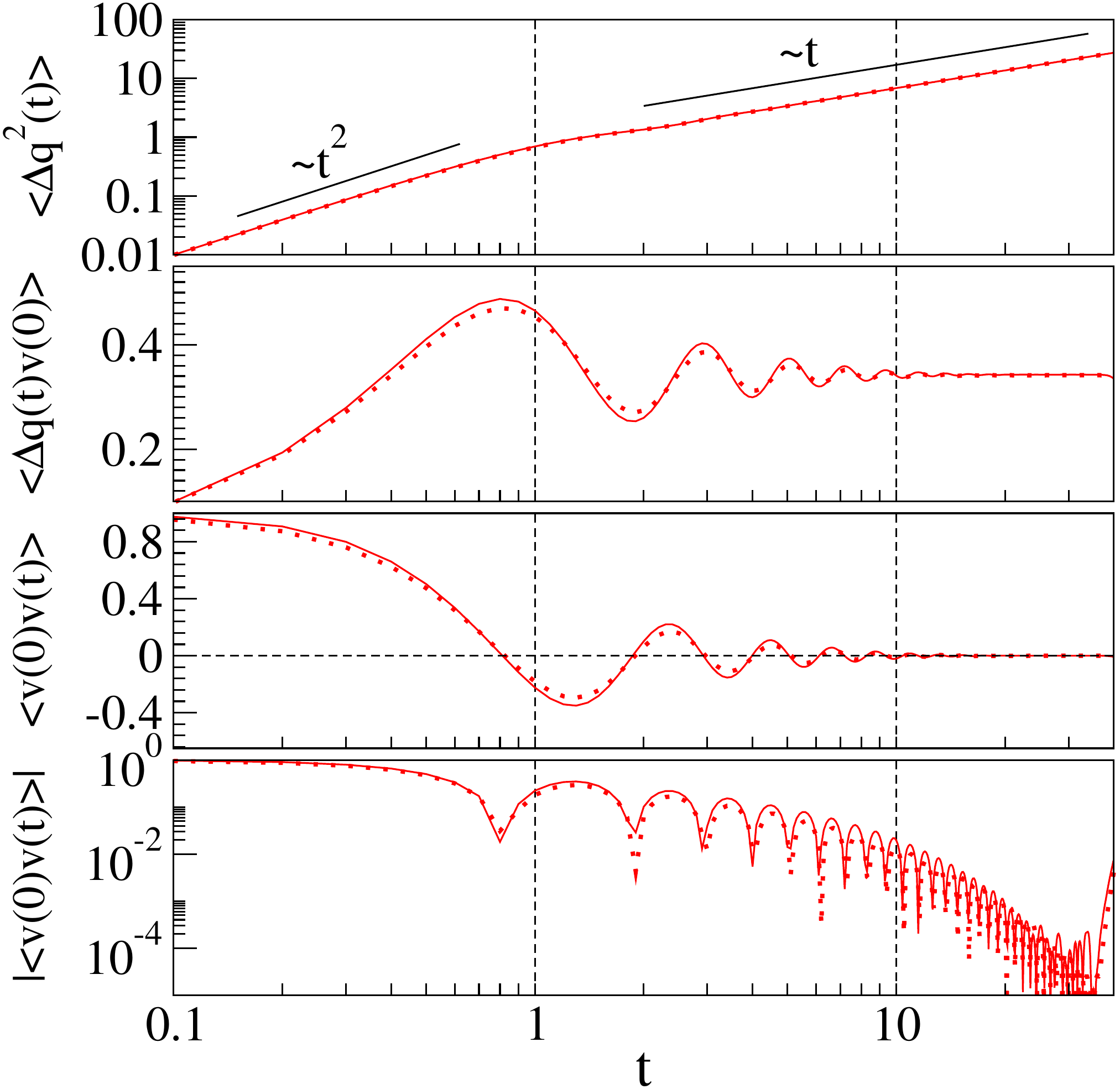}
\caption{Comparison of FPU chain with damped harmonic model: short time correlation functions of the central tagged particle
in a chain of size $N=65$ (solid line) compared with the predictions of the damped harmonic model (dashed line). 
The parameters of the chain here are same as in Fig.~(\ref{fpudat}). In this case $k_{\rm eff} = c^2 \approx 2.137$ and the only fitting parameter of the model is the damping constant which was fixed at $\gamma=0.1$.}
\label{fpudat3}
\end{figure} 

\begin{figure}
\includegraphics[scale=0.5]{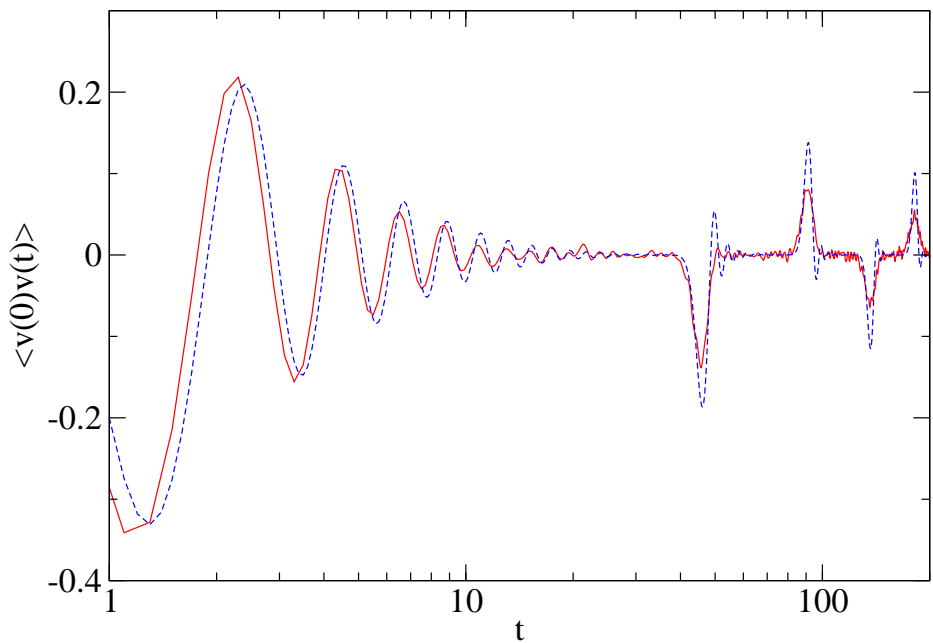}
\caption{Comparison of FPU chain with damped harmonic model: long time VAF of the central tagged particle in a chain of size $N=65$ (solid
line) compared with the predictions of the damped model (dashed line). The parameters of the chain here are again same as in Fig.~(\ref{fpudat}). The 
value of the spring constant again is $k_{\rm eff} = c^2 \approx 2.137$ and the damping constant was again set at $\gamma=0.1$.}
\label{fpudat4}
\end{figure}

Observing tagged particle dynamics constitutes a simple way of probing
the complex dynamics of an interacting many body system and has
been studied both
theoretically~\cite{roy13,jepsen65,lebowitz67,lebowitz72,percus10,evans79,kasper85,marro85,mazur60,bishop81,bagchi02,harris65,beijeren83,pincus78,beijeren91,rodenbeck98,kollman03,lizana08,gupta07,barkai09,barkai10,Krapivsky14,maiti10} 
and experimentally~\cite{hahn96,kukla96,wei00,lutz04,lin05,das10}. Much of the theoretical
studies on tagged particle diffusion have focused on one-dimensional
systems and discussed two situations where the microscopic particle
dynamics is (i) Hamiltonian~\cite{roy13,jepsen65,lebowitz67,lebowitz72,percus10,evans79,kasper85,marro85,mazur60,bishop81,bagchi02}
or (ii) stochastic~\cite{harris65,beijeren83,pincus78,beijeren91,rodenbeck98,kollman03,lizana08,gupta07,barkai09,barkai10,Krapivsky14}.
A hydrodynamic description of tagged particle diffusion has been
considered in \cite{pincus78,beijeren91}. Even for Hamiltonian
systems much of the work has been done on hard particle gases but
very little has been done on soft chains~\cite{mazur60,bishop81,bagchi02}.

Although there has been considerable work on transport properties of 
one-dimensional gases ~\cite{narayan02,beijeren12,dhar08}, this involves the
propagation of conserved quantities as a function of position and
time without reference to the identity of each particle. This changes
things considerably: for instance, conserved quantities propagate
ballistically for an equal mass hard particle gas, resulting in a
thermal conductivity proportional to $N,$ while tagged particle
dynamics in the same system is diffusive.  Thus, here we approach
the dynamics from a perspective that is different from the heat
conduction literature.

In this paper we present results for tagged particle correlations
for various one-dimensional systems.
In particular,
we compute the  mean-squared displacement (MSD) $\la [\Delta x(t)]^2 \ra $, the velocity auto-correlation function (VAF)  $\la v(t) v(0)\ra$, and 
$\la \Delta x(t) v(0) \ra$  of the central
particle, where $\Delta x(t)= x_M(t)-x_M(0)$ and $v(t)=v_M(t)$.
The average  $\la \cdots \ra$ is taken over initial configurations
chosen from the equilibrium distribution. (Details are given later
in the sections dealing with each system). Note that the three correlation
functions are related to each other as
\begin{eqnarray} 
\f{1}{2} \frac{d}{dt} \langle [\Delta x(t)]^2\rangle &=& 
\la \Delta x(t) v(t) \ra = \langle \Delta x(t) v(0)\rangle = D(t),\nonumber\\
\frac{d}{dt}\langle \Delta x(t) v(0)\rangle &=& \langle v(t)v(0)\rangle.
\label{diff_defn}
\end{eqnarray}
These results follow from $\Delta x(t) = \int_0^t v(t') dt'$ and 
$\langle v(t) v(t')\rangle = \langle v(t - t') v(0)\rangle;$ the last equation 
on the first line defines $D(t)$. 
We note that in a finite closed box of size $L$, $\langle [\Delta x(t)]^2\rangle$ is bounded and so $\lim_{t \to \infty} D(t)$ would always either vanish or oscillate. However one is usually interested in the nature of the MSD of the particle before it sees the effect of the boundary, and it is then appropriate to study the limit $D_\infty(t)=\lim_{L \to \infty} D(t)$.  
When the tagged particle shows normal diffusive behaviour, $D_\infty(t\rightarrow\infty)$
is a constant, which is the diffusion constant. On
the other hand, $D_\infty(t\rightarrow\infty)$ is zero for sub-diffusive, and
divergent for super-diffusive behaviour.

Here we examine the form of the correlations at both ``short times'', when boundary effects are not felt and 
hence the correlation functions are system size independent, and at ``long times'', after boundary effects show up. 
At short times the form of the correlation functions is un-affected by the boundary conditions, \emph{e.g.}  periodic or fixed boundary conditions. Hence, in simulations we sometimes use  periodic boundary conditions for short time studies, since one can then do an averaging over all particles, resulting in better statistics. The long time behaviour depends on boundary conditions and here we focus entirely on fixed or hard wall boundary conditions.
As we will see, system size effects typically show up at times $t \sim L/c$ where $L$ 
is the system size and $c$ the sound speed in the system. The short time regime typically has an initial ballistic regime, with $\la \Delta x^2(t)\ra \sim t^2$, 
and we will see that this is always followed by a diffusive regime, with $\la \Delta x^2(t)\ra \sim t$.

The rest of this paper is organized as follows.
In Section~\ref{sec:harm}, we present analytic results for the
harmonic chain, as an indicator of what one might expect for anharmonic
chains where an exact solution is not possible.  In Section~\ref{sec:FPU}, we  present simulation
results for the Fermi-Pasta-Ulam (FPU) chain of anharmonic oscillators, with a simple model of 
damped sound waves that compares well with the numerical results.  
In Section~\ref{sec:LJ}, we present the
simulation results for the Lennard-Jones (LJ) gas, and show that its correlation functions resemble 
those of  a hard-particle gas (obtained in Ref.~\cite{roy13}) at low densities, and an anharmonic chain at high densities.  
In Section~\ref{sec:hard}, we present simulation results for the alternate mass hard 
particle gas with large system sizes, and show that $D(t)$  saturates to a
constant in the large-$t$ limit, and that the approach to this asymptotic
limit is in agreement with the predictions of mode-coupling theory~\cite{beijeren}; 
this is in contradiction to the logarithmic decay for $D(t)$, that we had claimed in a  previous paper~\cite{roy13}, 
based on simulations on smaller system sizes.
Finally, in Section~\ref{sec:disc}, we provide a discussion and summary of our results.

\section{Harmonic chain}
\label{sec:harm}

We consider a harmonic chain of $N$ particles labeled $l=1,\ldots,N$.
The particles of masses $m$ are connected  by springs with stiffness
constant $k$. Let $\{ q_1,\ldots,q_N \}$ denote the displacements
of the particles about their equilibrium positions. The equilibrium
positions are assumed to be separated by a lattice spacing $a$ so
that the mass density is $ \rho=m/a$. We assume that the particles $l=0$ and $l=N+1$ are fixed so that $q_0=q_{N+1}=0$.  
The Hamiltonian of the system is
\bea
H= \sum_{l=1}^N \f{m}{2} \dot{q}_l^2 +\sum_{l=1}^{N+1} \f{k}{2} (q_l-q_{l-1})^2~. 
\label{harmH} 
\eea
Transforming to normal mode coordinates $q_l(t)= \sum_p
a_p(t) \phi_p(l)$ where
\bea
\phi_{p}(l)= \left[\f{2}{m(N+1)}\right]^{1/2} \sin (l p a)~~~ {\rm with}~~~
p =\f{n \pi}{(N+1)a}~,~~~~~n=1,\ldots,N
\label{norm_mode} 
\eea
brings  the Hamiltonian to the form
$H=\sum_p \dot{a}_p^2/2 + \omega_p^2 a_p^2/2$ with
\bea
\omega_p^2=2\f{k}{m}(1-\cos pa)~.
\label{norm_freq}
\eea
The normal mode equations of motion $\ddot{a}_p=-\omega_p^2 a_p$
are easily solved and lead to the following expression:
\bea
q_l(t)= \sum_{p} \phi_{p}(l)~ \left[ a_p(0)\cos \omega_p t  + 
\f{\sin \omega_p t}{\omega_p} \dot{a}_p(0)~\right]~.
\label{harmsoln}
\eea
We consider a chain in thermal equilibrium at temperature $T,$ i.e. 
$\la \dot{a}_p^2(0) \ra =\omega_p^2
\la a_p^2(0)\ra = k_B T$ and $\la\dot{a}_p(0) a_p(0)\ra = 0.$ For the middle 
particle, $l = M = (N + 1)/2$ (assuming odd $N$) and, since $p=n \pi/(N+1)a$,  therefore $\phi_p(l)\propto \sin(lpa)=\sin(n\pi/2)$ vanishes for even $n$. Defining $\Delta q(t) =
q_M(t) - q_M(0),$
\bea
\la [\Delta q(t)]^2 \ra = 
\f{ 8 k_B T}{m(N+1)} \sum_{n=1,3,\ldots} \f{\sin^2 (\omega_p t/2)}{\omega_p^2} ~.
\label{harmmsd}
\eea
 The correlations $\la  \Delta q(t) v(0) \ra$ and $\la v(t) v(0)\ra$ can be found by 
differentiating this expression, as in Eqs.(\ref{diff_defn}).

In Fig.~(\ref{harmdat})  and Fig.~(\ref{harm}) we 
plot the simulation results for the various correlation functions for different system sizes 
and find them to match the exact analytic results [Eq.(\ref{harmmsd}) and its derivatives].
As seen in Fig.~(\ref{harmdat}), an initial $\sim t^2$ growth in $\la\Delta q^2(t)\ra$ crosses over to a linear growth, indicating a diffusive regime that
is also seen in $\la\Delta q(t) v(0)\ra.$
After that, boundary effects set in and $\la (\Delta q)^2\ra/N$ is an almost periodic
function of $t/N$ (Fig.~(\ref{harm})). This is
somewhat surprising since we are averaging over an initial equilibrium
ensemble with all normal modes, and $\omega_p \approx c p$ (where $c = a \sqrt{k/m}$ is the wave speed)
only for small $p.$

The behaviour of $\la [\Delta q(t)]^2 \ra$ for the harmonic chain can be understood
in detail by analyzing the different time regimes of Eq.~(\ref{harmmsd})~.
There are three regimes of $t$ to consider:

(i) When $\omega_N t << 1,$ $\sin^2(\omega_n t/2) \approx \omega_n^2
t^2/4.$ (We use $\omega_n$ to denote the normal mode frequencies in Eq.(\ref{norm_freq}), but with $p = n\pi/[(N+1) a].$)
The right hand side of Eq.~(\ref{harmmsd}) is then equal
to $k_B T t^2/m$~. This approximation is valid as long as $\omega_N
t = 2 c t/a$ is small.

(ii) In the second regime, $\omega_N t >> 1 >> \omega_1 t,$ and the
sum can be replaced by an integral:
\begin{equation}
\frac{ 4 k_B T}{m(N+1)}\int_{1}^N dn \frac{\sin^2(\omega_n t/2)}{\omega_n^2} 
\approx
\frac{2 k_B T a t }{\pi m c}
\int_0^\infty dy
\frac{\sin^2(y)}{y^2\sqrt{1 - (a y/c t)^2}} ~
\label{harm1}
\end{equation}
where we have changed variables from $n$ to $y =  \omega_n t/2$ and used $\omega_N t >> 1 >> \omega_1 t$
to change the limits of the integral. Expanding $[1 - (ay/ct)^2]^{-1/2}$ in a binomial series and keeping
only the first term of the expansion, we get the leading
order behaviour in t as $ \la [\Delta q(t)]^2 \ra = (k_B
T t/\rho c)$.  The linear $t$-dependence  implies diffusive behaviour
with a diffusion constant $D=k_B T/(2 \rho c)$.

The velocity auto-correlation
function in this regime can be obtained by differentiating the first expression in Eq.(\ref{harm1}), 
as in Eqs.(\ref{diff_defn}): 
\begin{equation}
\la v(t) v(0)\ra = \frac{ k_B T}{m(N+1)}\int_{1}^N dn \cos(\omega_n t)~.
\label{harm2}
\end{equation}
Substituting $z = \omega_n/(2 c/a) = \sin(n\pi/2 (N + 1)),$ this is equivalent to 
\begin{equation}
\la v(t) v(0) \ra  = 
\frac{ 2 k_B T}{\pi m}\int_0^1 dz \frac{\cos(2 c t z/a)}{\sqrt{1-z^2}} = \frac{ k_B T}{m} J_0(2 c t/a)~\sim \frac{\cos(2 c t/a - \pi /4)}{t^{1/2}}~~{\rm for}~ t\rightarrow\infty,
\end{equation}
using the asymptotic properties of Bessel functions~\cite{mazur60}. 
This is shown in the lower panels of Fig.~\ref{harmdat}.

(iii) In the third regime, the results depend on boundary conditions and our analysis here is for fixed boundary conditions. In this case $\omega_1 t $ is no longer small. As a first
approximation, we set $\omega_n $ to be equal to $c n\pi/[(N+1) a] \approx c n \pi/Na.$ Then
Eq.~(\ref{harmmsd}) becomes
\begin{equation}
\langle [\Delta q(t)]^2 \rangle  = 
\frac{8 k_B T N a}{\rho c^2 \pi^2} \sum_{n=1,3,5\ldots} \frac{1}{n^2} \sin^2 \left(\f{c n \pi  t}{2N a}\right)~.
\end{equation}
This is a periodic function in $t$ with a period of $2 N a/c.$
Since the sum is dominated by $n << N,$ we see that our first approximation is a reasonable one.
More accurately, we expand $\omega_n$ to one order higher:
\begin{equation}
\omega_n = \frac{c n \pi}{N a} [1 - n^2 \pi^2/(24 N^2) + \cdots]
\end{equation}
and evaluate the sum at $t = 2 j N a/c.$ We have 
\begin{equation}
\langle [\Delta q(t)]^2 \rangle  = 
\frac{8 k_B T N a}{\rho c^2\pi^2} \sum_{n = 1,3,\ldots} \frac{1}{n^2 [1 - O(n/N)^2]}
\sin^2 (j n^3 \pi^3/(24 N^2)).
\end{equation}
Approximating the sum by an integral and changing
variables to $y = n j^{1/3}/N^{2/3}$, we get
\begin{equation}
\langle [\Delta q(t)]^2 \rangle  = 
\frac{8 k_B T (N j)^{1/3} a}{\rho c^2 \pi^2}\int_{O(N^{-2/3})}^{O(N^{1/3})}
\frac{\sin^2 (y^3\pi^3/24)}{y^2[1 - O(y^2/(jN)^{2/3})]} dy.
\end{equation}
As $N\rightarrow\infty,$ the integral converges to an $N$-independent
value of $0.8046\ldots,$ so that the function is $O(N j)^{1/3}.$
We note that this is small compared to the
$O(N)$ value of the function at its maxima, but that it increases
steadily with $j,$ as expected in a dispersive system. In a more careful analysis, the locations of 
the minima are taken to be $2 j N a/c + \delta_j$ and the $\delta_j$'s evaluated to 
leading order, but this does not change the fact that the minima are $O(N j)^{1/3}.$ A 
similar analysis shows that the function at its 
maxima is equal to $k_B T N a/(\rho c^2) - O(j N)^{1/3}.$

\section{Fermi-Pasta-Ulam chain}
\label{sec:FPU}

We now turn to a numerical study of tagged particle motion in a
chain of nonlinear oscillators.  The Hamiltonian of the Fermi-Pasta-Ulam
 chain we study is taken to be:
\bea
H= \sum_{l=1}^N \f{m}{2} \dot{q}_l^2 +\sum_{l=1}^{N+1} \left[ \f{k}{2} (q_l-q_{l-1})^2~+ \f{\nu}{4} (q_l-q_{l-1})^4 \right],\label{fpuH}
\eea
where the $q_l$'s are the displacements from equilibrium positions. We fix
the particles at the boundaries by setting $q_0=0$ and $q_{N+1}=0$.
The corresponding equations of motion are:
\begin{equation} 
m \ddot q_l = -k (2 q_l - q_{l+1} - q_{l-1})
- \nu (q_{l} -q_{l-1})^3 -\nu (q_{l} -q_{l+1})^3 ~ .\label{fpueqm}
\end{equation}
In this case there is no analytic solution of the equations of
motion and we evaluate $\la [\Delta q(t)]^2 \ra$ and other
correlation functions for the central tagged particle through direct
MD simulations. We prepare the initial thermal equilibrium state
by connecting all the N particles to Langevin-type heat baths and
evolving the system for some time.  The heat baths are then removed
and, starting from the thermal initial condition, the system is
evolved with the dynamics of Eq.~(\ref{fpueqm}). The average $\la
...\ra$ is obtained by creating a large number of independent thermal
initial conditions. 

The simulation results are plotted in Fig.~(\ref{fpudat}) and Fig.~(\ref{fpudat2}).
Comparing to the analogous figures for the harmonic chain, we see that the plots have some  similarity, as well as significant differences. At short times [Fig.~(\ref{fpudat})], 
there is again a crossover from ballistic [$\la\Delta q^2(t)\ra \sim t^2$] to diffusive 
[$\la\Delta q^2(t)\ra \sim t$] behaviour. Moreover, the velocity auto-correlation
function $\la v(0) v(t)\rangle$ shows oscillatory behavior as in the harmonic case, however the  
damping is much faster (exponential) for the FPU chain than for the harmonic chain (power-law decay).
Also, at long times [Fig.~(\ref{fpudat2})], we can see $\la \Delta q^2(t)\ra$ converging to the expected 
equilibrium values, for small system sizes, unlike the harmonic chain where the oscillations persist for ever. For larger system sizes, the scaling collapse seen for the harmonic chain is not obtained.

As a simplified model of this system, we consider an effective harmonic description of the system.   The  nonlinear terms in Eq.(\ref{fpueqm}) normally couple the normal modes of the linear system, and we now assume that they  
can be replaced by momentum conserving dissipation and noise terms --- thus for
any mode, all the other modes act as a heat bath.  It is important to add {\emph{momentum conserving}} noise and dissipation since these are generated internally from the systems dynamics and have to preserve the conservation laws. Note that this is different from the effective harmonization technique of Refn.~\cite{lizana10}, where the original dynamics is stochastic, and the noise and dissipation  already exists. 
Our effective harmonic model with noise and dissipation is described by the following equations of motion:
\bea
m \ddot{q}_l=-k_{\rm eff} (2 q_l-q_{l+1}-q_{l-1})-\gamma(2 \dot{q}_l-\dot{q}_{l+1}-\dot{q}_{l-1})+(2 \xi_l-\xi_{l+1}-\xi_{l-1})~,
\label{efeq}
\eea  
where $k_{\rm eff}$ is an effective spring constant, $\gamma$ a damping constant and $\xi_l$ are noise terms whose properties will be specified later. We consider  fixed boundaries condition $q_{0}=0$ and $q_{N+1}=0$.  As for
the harmonic oscillator, we transform to normal mode coordinates defined
in Eq.(\ref{norm_mode}) [$\phi_{p}(l)= \sqrt{2}/\sqrt{m(N+1)} \sin (l p a)~~~ {\rm with}~~~
p a =n \pi/(N+1)~,~n=1,\ldots,N$], with $\xi_l(t) = \sum_p \tilde\xi_p(t)\phi_p(l).$
The normal mode coordinates $a_p(t)$ now satisfy the equation of motion
\bea
 \ddot{a}_p(t)+  \omega_p^2 a_p(t)&=& -\frac{\gamma}{k_{\rm eff}} \omega_p^2 \dot{a}_p(t) + \frac{\omega_p^2}{k_{eff}} \tilde{\xi}_p(t)~, \nn\\
\text{where} \quad \omega_p^2 &=& \frac{2k_{\rm eff}}{m}(1-\cos pa). \nn
\eea
To ensure equilibration of the modes
we choose Gaussian noise with zero mean and two point correlations
given by
\bea
\la \tilde{\xi}_p(t)~\tilde{\xi}_{p'}(t') \ra = \f{2 k_{\rm eff}\gamma k_B T}{m\omega_p^2} \delta (t-t') ~\delta_{p,p'}~. \nn
\eea
In steady state, $a_p(t) = \omega_p^2 \int_{-\infty}^t G(t - t') \tilde\xi_p(t') dt',$ where $G(t - t')$ is 
the Green's function for the equation of motion. 
After some straightforward computations we finally get
\begin{subequations}
\label{hydeqsol}
\bea
\la q_l(t) q_l(0)\ra &=& k_B T \sum_p \f{\phi_p^2(l)}{\omega_p^2} e^{-\alpha_p t} \left[ \cos (\beta_p t) +\f{\alpha_p}{\beta_p} \sin (\beta_p t) \right], \\
\la {q}_l(t) v_l(0)\ra &=& k_B T \sum_p \f{\phi_p^2(l)}{\beta_p} e^{-\alpha_p t} \sin (\beta_p t)~,  \\
\la v_l(t) v_l(0)\ra &=& k_B T \sum_p \phi_p^2(l) e^{-\alpha_p t} \left[ \cos (\beta_p t) -\f{\alpha_p}{\beta_p} \sin (\beta_p t) \right]~,  \\
{\rm where}~~\alpha_p &=&\f{\gamma \omega_p^2}{2 k_{\rm eff}}~,~~~\beta_p=(-\alpha_p^2+\omega^2_p)^{1/2}~.\nn
\eea
\end{subequations}
Taking $N$ to be odd, we get for the middle particle $l=(N+1)/2$
\begin{subequations}
\label{hyd}
\bea
\la q(t) q(0)\ra &=& \f{2k_B T}{m(N+1)} \sum_{n=1,3,\ldots}  \f{e^{-\alpha_p t}}{\omega_p^2} \left[ \cos (\beta_p t) +\f{\alpha_p}{\beta_p} \sin (\beta_p t) \right]~, \\
\la {q}(t) v(0)\ra &=& \f{2 k_B T}{m(N+1)} \sum_{n=1,3,\ldots} \f{e^{-\alpha_p t}}{\beta_p} \sin (\beta_p t)~, \\
\la v(t) v(0)\ra &=& \f{2k_B T}{m(N+1)} \sum_{n=1,3,\ldots}  e^{-\alpha_p t} \left[ \cos (\beta_p t) -\f{\alpha_p}{\beta_p} \sin (\beta_p t) \right]~, 
\eea
\end{subequations}
If we take the $t \to \infty$ limit first in the middle equation in Eq. \ref{hyd} then we get $D=0$. As discussed in the Sec.~(\ref{sec:intro})  we need to take the $N \to \infty$ limit before $t \to \infty$ limit, thereby exploring the infinite system diffusive behaviour.
In the limit $N \to \infty$, the above equations give:
\begin{subequations}
\label{hydNoo}
\bea
\la q(t) q(0)\ra &=& \f{k_B T a}{m \pi } \int_0^{\pi/a} dp  \f{e^{-\alpha_p t}}{\omega_p^2} \left[ \cos (\beta_p t) +\f{\alpha_p}{\beta_p} \sin (\beta_p t) \right]~, \\
\la {q}(t) v(0)\ra &=& \f{k_B T a}{m\pi} \int_0^{\pi/a} dp \f{e^{-\alpha_p t}}{\beta_p} \sin (\beta_p t)~,\\
\la v(t) v(0)\ra &=& \f{k_B T a}{m \pi } \int_0^{\pi/a} dp  e^{-\alpha_p t} \left[ \cos (\beta_p t) -\f{\alpha_p}{\beta_p} \sin (\beta_p t) \right]~.
\eea
\end{subequations}
The diffusion constant is obtained as
\bea
\lim_{t \to \infty} \la q(t) v(0)\ra = \f{k_B T a}{m \pi c} \int_0^\infty dx \f{\sin ( x )}{x } = \f{k_B T}{2 \rho c} \label{hydD}
\eea
where, as before, $\rho$ is the mass per unit length and $c$ is the
speed of sound. The long time behaviour of the VAF in Eq.~(\ref{hydNoo}) can be estimated and we find $\la v(t) v(0) \ra \sim \exp(-\gamma t/m) \sin (2 c t/a)$. 

Our  effective damped harmonic theory  in fact follows from the full  
hydrodynamic equations discussed in Ref. \cite{spohn13}. Here we briefly outline such a derivation. In  the hydrodynamic theory, one starts with equations for the three coarse-grained conserved fields, corresponding to the extension $r_l=q_{l+1}-q_l$, momentum $p_l$ and  energy $e_l$. These are given by
\begin{align}
\f{\p r(x,t)}{\p t}&=\f{\p p(x,t)}{\p x}~, \nn \\
\f{\p p(x,t)}{\p t}&=-\f{\p P(r,\hat{e})}{\p x}~, \nn \\
\f{\p e(x,t)}{\p t}&=-\f{\p P(r,\hat{e}) p(x,t)}{\p x}~,
\label{hydeq}
\end{align}
where $P(r,\hat{e})$ is the local pressure and is a function of the local extension and energy $\hat{e}=e-p^2/2$. In the above equations we have made the identification $l \to x$ and $\p/\p x$ can be thought of as a discrete derivative. 
Next, one looks at fluctuations of the conserved fields about their equilibrium value and define the fields $u_1=\delta r= r-\la r\ra$, $u_2=p$ and $u_3=\delta e= e-\la e\ra$, where the angular brackets denote equilibrium averages. Then, expanding the pressure around its equilibrium value, $P_{eq}$, we get from Eq.~(\ref{hydeq}) the following linear equations
\begin{align}
\f{\p \delta r(x,t)}{\p t}&=\f{\p p(x,t)}{\p x}~, \nn \\
\f{\p p(x,t)}{\p t}&=-\f{\p P_{eq}(r,\hat{e})}{\p r} \f{\p \delta r(x,t)}{\p x} -\f{\p P_{eq}(r,\hat{e})}{\p e} \f{\p \delta e(x,t)}{\p x}, \nn \\
\f{\p \delta e(x,t)}{\p t}&=-P_{eq} \f{\p p(x,t)}{\p x}~.
\label{linhydeq}
\end{align}
Substituting for $\p p(x,t)/ \p x$ from the first line into the third line of Eq.~(\ref{linhydeq}),  and operating $\p/\p x$ on both sides, we get $\p \delta e(x,t)/\p x = -P_{eq}\p \delta r(x,t)/\p x$. Substituting this in the second line above, we obtain
\begin{align}
\f{\p p(x,t)}{\p t} &=  \left[-\f{\p P_{eq}(r,\hat{e})}{\p r} + P_{eq}(r,\hat{e})\f{\p P_{eq}(r,\hat{e})}{\p e}\right]\f{\p \delta r(x,t)}{\p x}. 
\label{linhyd}
\end{align}
The terms within brackets is precisly the expression for $c^2$ given  in \cite{spohn13}, where an expression for it in the constant temperature/pressure ensemble has been given. We now observe that, on adding the momentum conserving noise and damping terms, Eq.~(\ref{linhyd})  is equivalent to our Eq.~(\ref{efeq}), upon identifying $c^2=k_{\rm eff}/m$.
 
Using the  expression for sound speed in \cite{spohn13}, for our symmetric FPU chain,  gives  $c=a [k_B T/(m \la r^2 \ra)]^{1/2}$.
For our simulation parameters, we get $c = 1.46\ldots$ and hence from Eq.~(\ref{hydD}), $D_{\text{FPU}} = 0.342\ldots$, which is in excellent agreement with the simulation results [Fig.~(\ref{fpudat})].
In Fig.~(\ref{fpudat3}) and Fig.~(\ref{fpudat4}) we compare the predictions
of Eq.~(\ref{hyd}) with the simulation results of the FPU chain of Fig.~(\ref{fpudat}) of size $N=65$.  
The effective spring constant $k_{\rm eff}$ is obtained from the $c$ above and $\gamma$ is the only fitting
parameter used. We see that this model seems to provide a good description of tagged particle diffusion in this system. We have also performed some numerical simulations of the asymmetric FPU chain, where, we find that the qualitative features 
of the various correlation functions remain the same. 
At a quantitative level we find less agreement with the effective damped harmonic  model, for example the diffusion constant differs from the prediction in Eq.~(\ref{hydD}). This could be related to the stronger effect of nonlinearity in the case of the asymmetric chain.

\begin{figure}
\includegraphics[scale=0.55,angle=0]{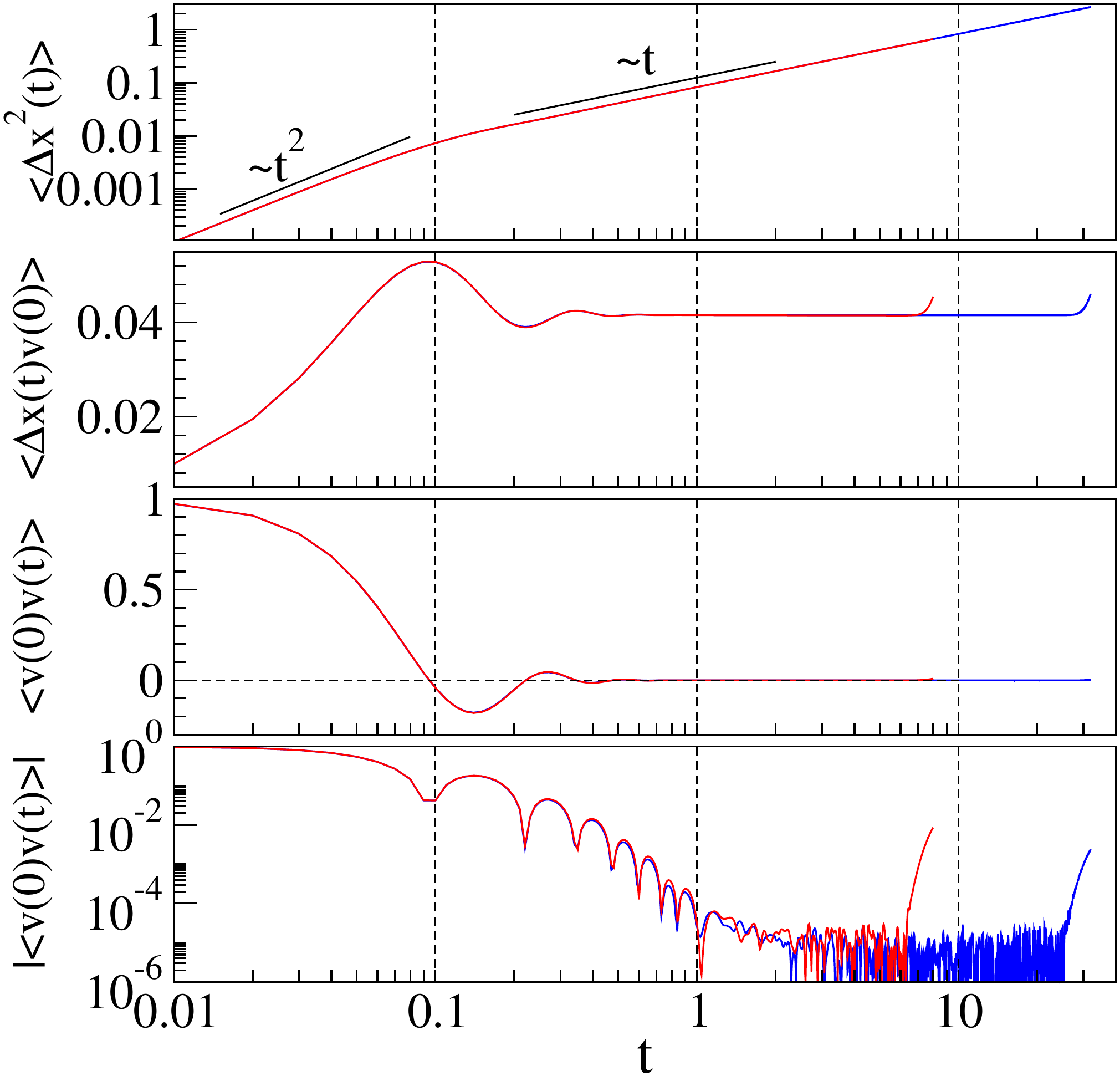}
\caption{LJ chain (high density): short time correlation functions of the central tagged particle
in systems of sizes $N=129$ (red) and $N=513$ (blue) with inter-particle
separation $1.0$ and $k_BT=1$. All the particles are of mass $1.0$. Periodic boundary conditions were used here. }
\label{lj1}
\end{figure} 

\begin{figure}
\includegraphics[scale=0.55,angle=0]{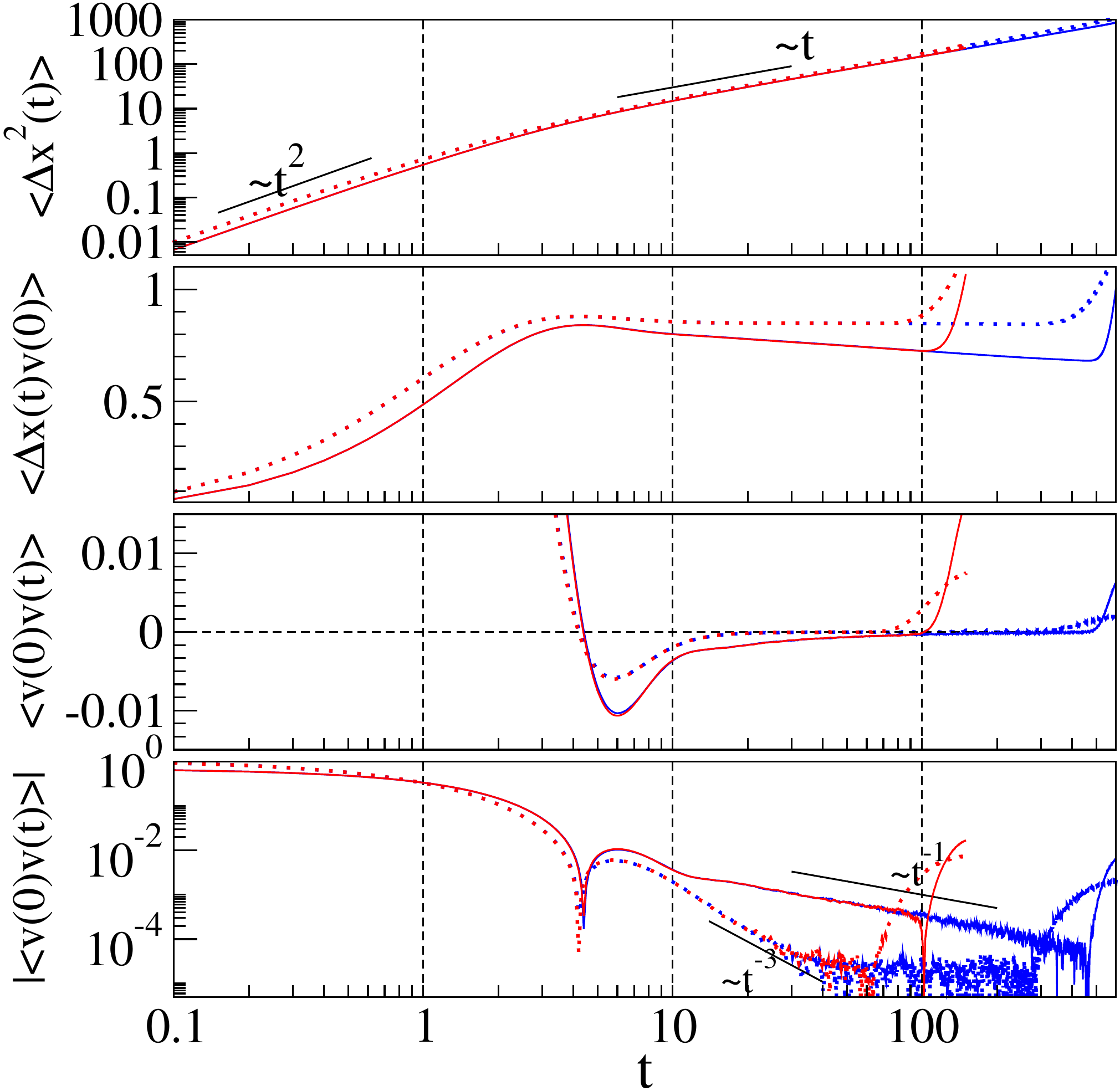}
\caption{LJ chain (low density): short time correlation functions of the central tagged particle
in systems of sizes $N=129$ (red) and $N=513$ (blue) with inter-particle
separation $3.0$ and $k_BT=1$.  The equal mass case (with $m=1.0$) is represented by dotted lines,
while solid lines represent the alternate mass case (with masses $1.5$ and $0.5$ alternately). 
Periodic boundary conditions were used here.}
\label{lj2}
\end{figure} 

\begin{figure}
\includegraphics[scale=0.3]{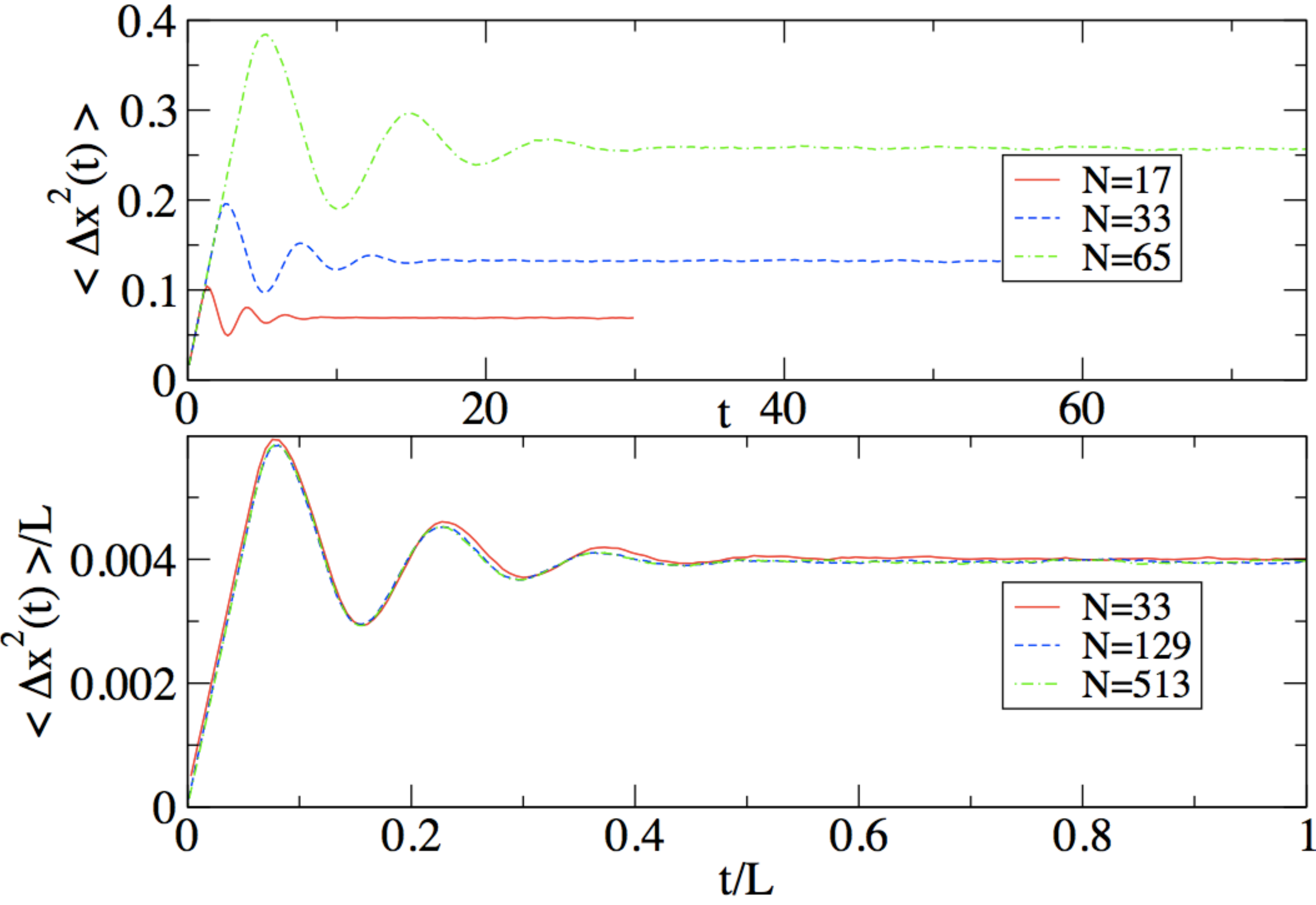}
\caption{LJ chain (high density): long time MSD of the central tagged particle in systems 
of different sizes $N$ with inter-particle separation $1.0$, $k_BT=1$ and fixed boundary conditions. 
All the particles are of mass $1.0$. The sound speed in this case, as calculated within the effective model, is $c \approx 13.02$.}
\label{ljdat}
\end{figure}

\begin{figure}
\includegraphics[scale=0.55]{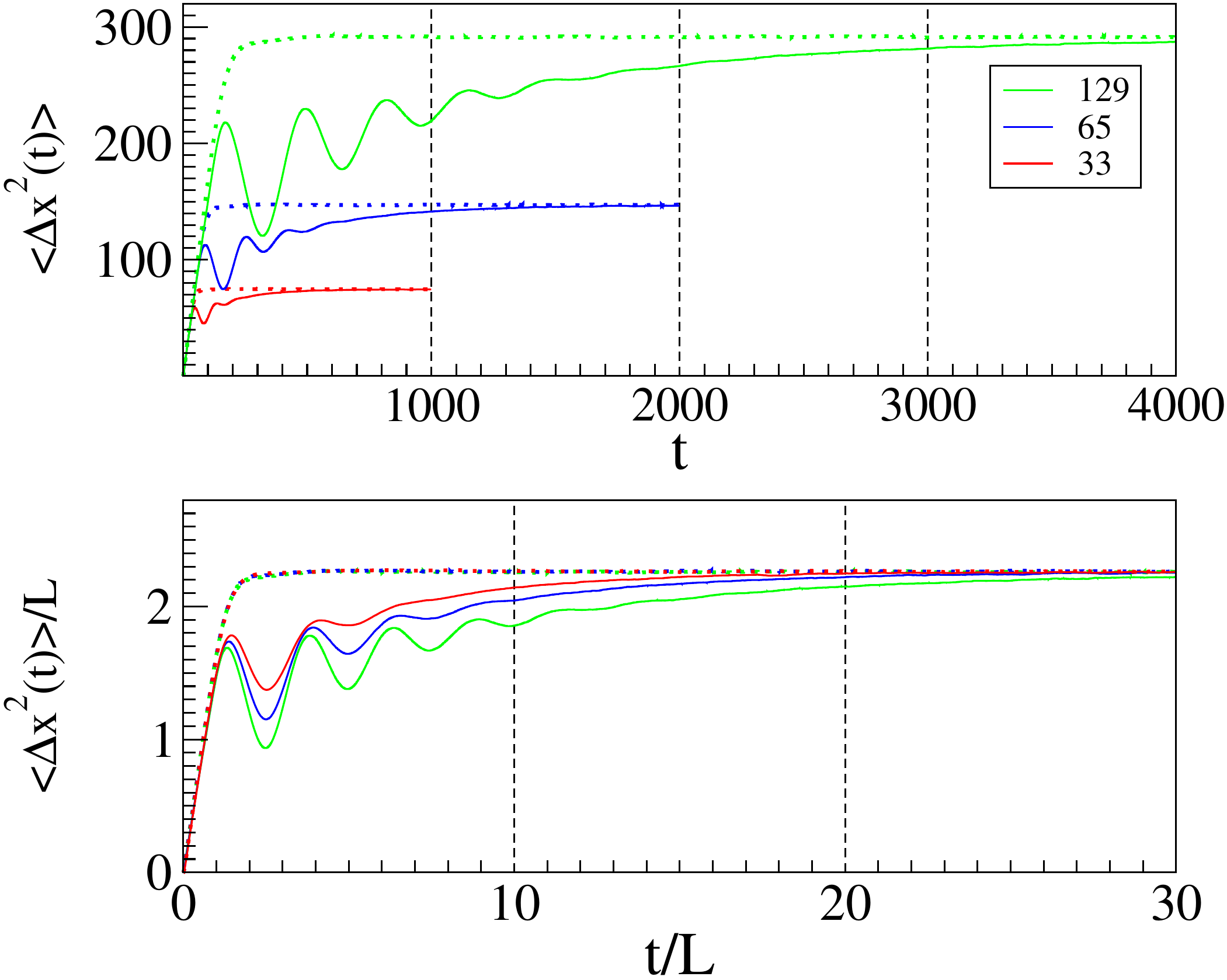}
\caption{LJ chain (low density): long time MSD of the central tagged particle in systems
of different sizes $N$ with inter-particle separation $3.0$, $k_BT=1$ and  fixed boundary conditions. 
The equal mass case (with $m=1.0$) is represented by dotted lines,
while solid lines represent the alternate mass case (with masses $1.5$ and $0.5$ alternately). The sound speed for the equal mass case, as calculated within the effective model, is $c\approx 0.8$. }
\label{ljdat3p0}
\end{figure}

\begin{figure}
\includegraphics[scale=0.55,angle=0]{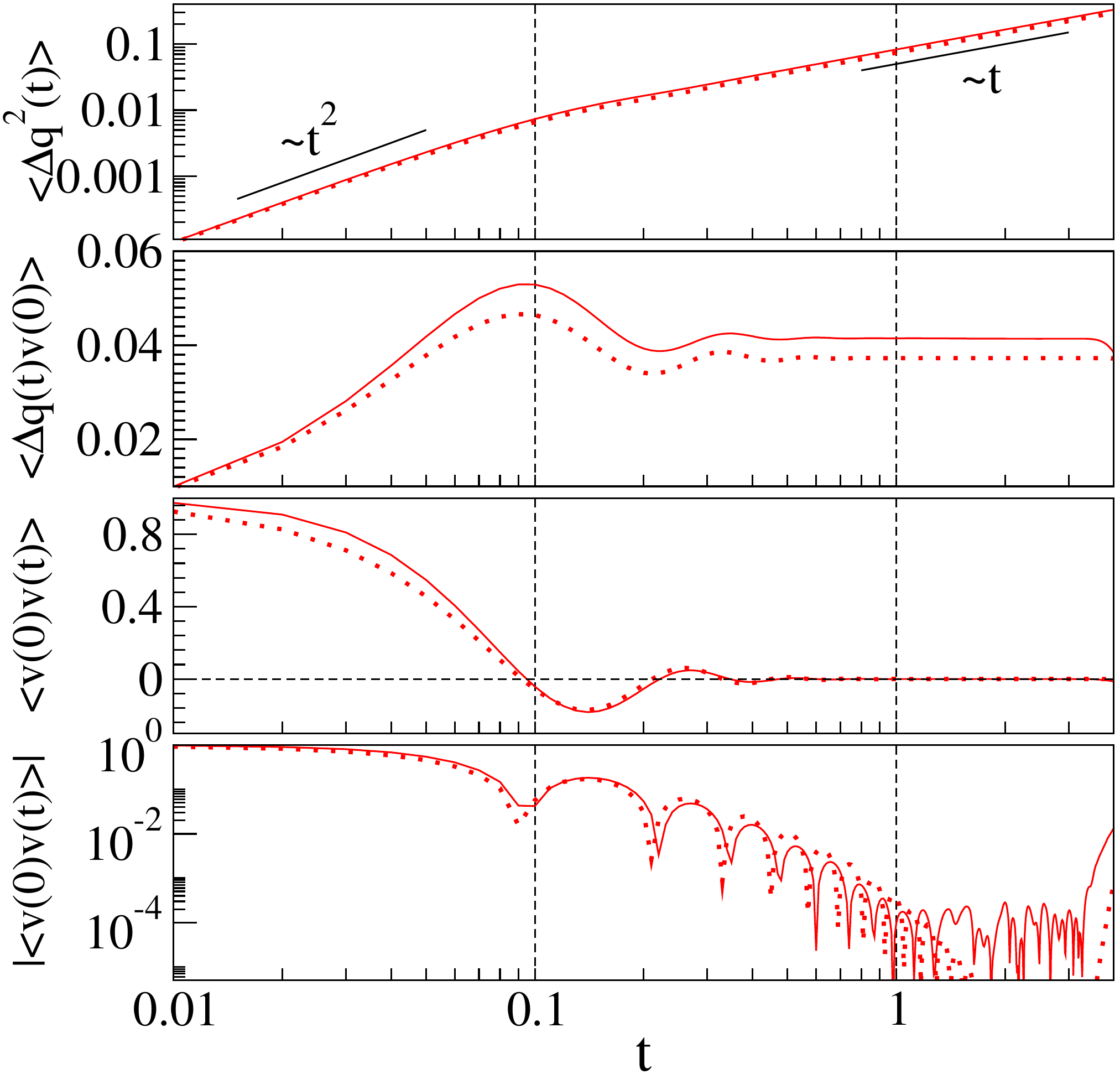}
\caption{Comparison of LJ chain with effective damped harmonic model: short time correlation functions of the central tagged particle
in chain of size $N=65$, with inter-particle separation $1.0$, $k_BT=1$ and $m=1.0$ (solid line), compared with the predictions
of the damped harmonic model (dashed line). In this case $k_{\rm eff}=c^2 \approx 169.767$ and the only fitting parameter
of the model is the dissipation constant $\gamma$ which was fixed at $\gamma=0.3$.}
\label{lj1stcomp}
\end{figure}

\begin{figure}
\includegraphics[scale=0.5,angle=0]{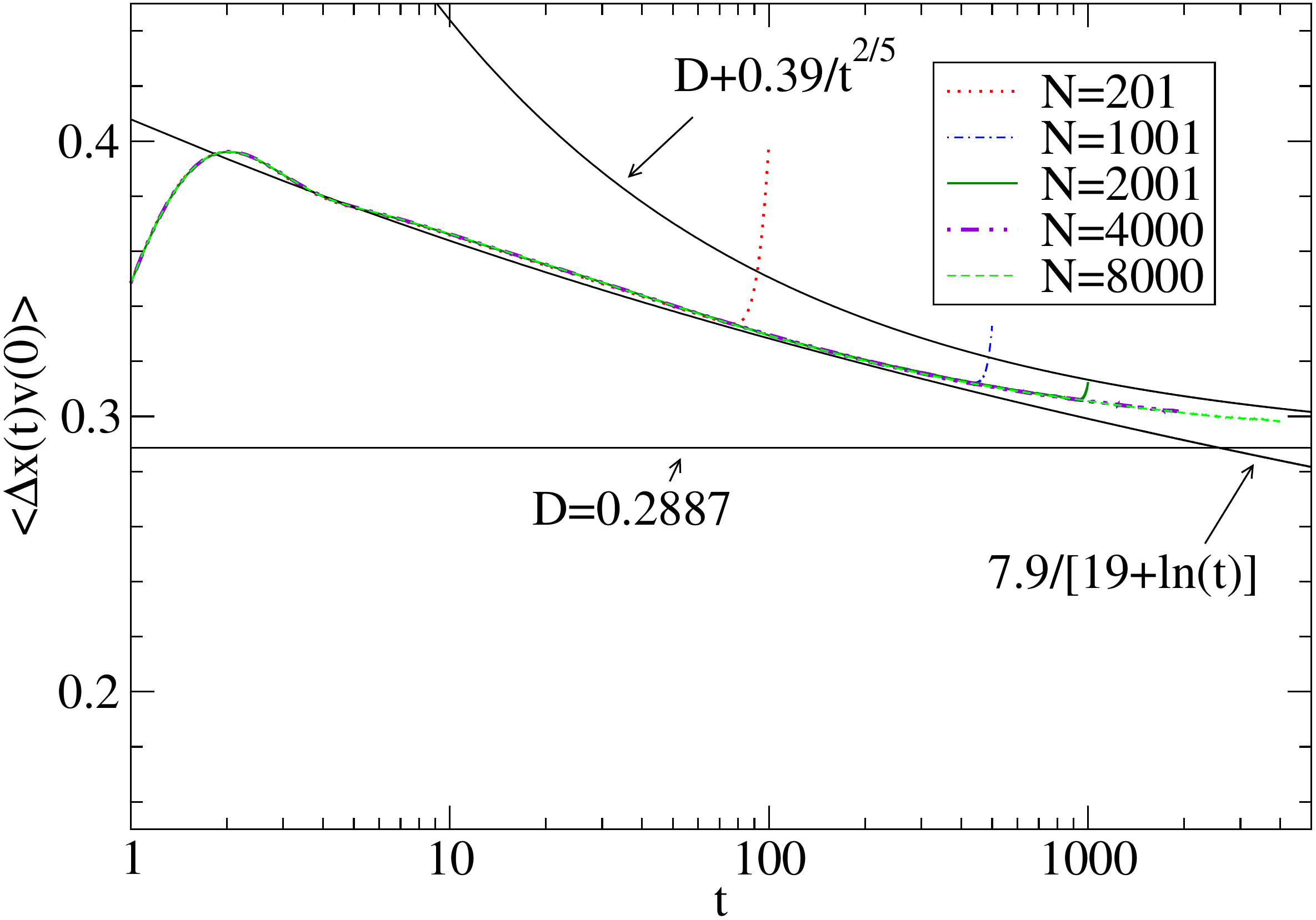}
\caption{Alternate mass hard particle gas: plot of $D(t) = \la \Delta x(t) v(0)\ra$ for systems of different sizes. The solid line with a
logarithmic decay was used as a fitting function in Ref.~\cite{roy13},
and is seen to
deviate from the largest system size data. The other solid line is
the prediction of mode-coupling theory~\cite{beijeren} and gives  a slow
power-law decay to the asymptotic diffusion constant $D=k_B T/(2\rho c) = 0.2886\ldots$.
In these simulations we used periodic boundary conditions since these give better statistics. 
Here the masses are $1.5$ and $0.5$ alternately and we have taken $k_BT=1$ and $\rho=1$.}
\label{altm}
\end{figure}

\section{Lennard-Jones gas} 
\label{sec:LJ}

The mean-squared displacement $\langle \Delta q^2(t)\rangle$ for the
FPU (and harmonic) chain seems to have a similar dependence on $t$
as for a hard particle gas~\cite{roy13}, with an initial $\sim t^2$
increase crossing over to a $\sim t$ dependence (before sound waves are reflected from boundaries, or, in the $N\to \infty$ limit). However, derivatives
of this correlation function, $\langle \Delta q(t) v(0)\rangle$ and
$\la v(0) v(t)\ra$ show differences. For the FPU chain, $\la \Delta
q(t) v(0)\ra$ approaches a constant rapidly. Although this is less
rapid for a harmonic chain, it is nevertheless clear that the large
$t$ limit is a constant. On the other hand, for the alternate mass hard 
particle gas, $\la \Delta x(t) v(0)\ra$ decreases slowly as $t$ increases,
with a levelling off  at very long times \cite{roy13} [see also Sec.~(\ref{sec:hard})]. 
Turning to the velocity auto-correlation function, for the FPU chain this 
has a damped oscillatory behaviour, while there are no --- or overdamped ---
oscillations in the hard particle velocity auto-correlation function.

The hard particle gas may be considered as an extreme case of a non-linear 
oscillator chain, but it is a singular limit of this family. To see if 
the differences between the correlation functions for the two cases are significant, we study the Lennard-Jones gas.
The Hamiltonian of the Lennard-Jones gas  is taken to be
\bea
H= \sum_{l=1}^N \f{m}{2} \dot{x}_l^2 +\sum_{l=1}^{N+1} \left[ \f{1}{(x_l-x_{l-1})^{12}}~-~ \f{1}{(x_l-x_{l-1})^6}\right] \label{ljH} 
\eea
where $x$'s are the positions of the particles.  At low densities,
one would expect the particles to behave approximately like free
particles, with a repulsive force between neighbouring particles
when they come close to each other. Since the repulsion occurs over
a distance that is small compared to the mean inter-particle separation,
the system is similar to a hard particle gas. On the other hand,
at high densities, the particles should remain close to their
equilibrium positions with small deviations, resulting in behaviour
more like the FPU chain (with both cubic and quartic anharmonic terms).

As for the FPU
chain we evaluate the correlation functions of the central particle
from molecular dynamics simulations.  The particles are inside a
box of length $L$ and we fix particles at the boundaries by setting
$x_0=0$ and $x_{N+1}=L$. The mean inter-particle spacing is thus
$a=L/(N+1)$. The simulation results are given in Fig.~(\ref{lj1}) and
Fig.~(\ref{lj2}) for short times and Fig.~(\ref{ljdat}) and Fig.~(\ref{ljdat3p0}) 
for long times. In these simulations we have taken $k_B T = 1$.

As expected, we observe in Fig.~(\ref{lj1}) and
Fig.~(\ref{ljdat}) that at high density the behaviour
is similar to that of the FPU chain. At low densities, Fig.~(\ref{lj2}) and Fig.~(\ref{ljdat3p0}),
the behaviour resembles that of the hard particle gas. The figures correspond to $a=1.0$ and $a=3.0$ respectively. 
The  $-1/t$ decay of the VAF in Fig.~(\ref{lj2}) is similar to that observed in \cite{roy13} for the alternate mass hard particle gas. However this behaviour cannot persist since it would give a negative (and infinite) diffusion constant.  We expect that, as for the alternate mass hard particle gas (discussed briefly in the next section), at long times this will change to  a decay of the form
$~-1/t^{7/5}$. Verifying this numerically appears quite challenging.

The effective damped harmonic model used for the FPU chains works reasonably well for the correlation functions of the high density Lennard-Jones gas (Fig.~(\ref{lj1stcomp})) but the quantitative agreement, for example with the predicted 
diffusion constant, is not very good. We believe that this is possibly because of the cubic non-linearity (in an  FPU description valid for small displacements) being significant.  
In the low density regime where the system behaves like a hard particle gas, the effective harmonic model does not work at all (figure not shown). 
This is because the model still has inherent oscillations in the velocity auto-correlation function which are absent for a hard particle gas.

\section{Hard particle gas}
\label{sec:hard}

Finally, we present simulation results for the alternate mass hard
particle gas that extend our results in Ref.~\cite{roy13}. In that
paper, results for various tagged particle correlation functions in the hard particle gas (equal and alternate mass cases) were presented.
Here, we limit ourselves to a more extensive study of $D(t)$  for the 
alternate mass hard particle gas. In 
Ref.~\cite{roy13} we studied the hard particle gas for system sizes up to 801; 
the slow decay of $\la \Delta x(t)v(0)\ra$ as a function of time 
led us to conclude that the system
is subdiffusive, with $D(t)\sim a /(b + \ln t).$ Here we present
simulation results for much larger system sizes in Fig.~(\ref{altm}).
We find a deviation at long times from our earlier conclusion, and
see that the numerical results are consistent with the prediction
from mode coupling theory~\cite{beijeren} of $\la \Delta x(t)
v(0) \ra = 0.2887 + 0.39/t^{2/5}.$ This implies a $-1/t^{7/5}$ decay form for the VAF. In particular, this indicates
that the system is diffusive, contrary to our earlier conclusion in \cite{roy13}.

\section{Discussions}
\label{sec:disc}

In this paper, we have studied the motion of the middle tagged particle in a one-dimensional chain of $N$ particles, evolving with Hamiltonian dynamics with nearest neighbour interaction potentials. We examined the form of various correlation functions both in the regime of short times (corresponding to infinite system sizes) and long times (to see finite size effects). 
We find that, both harmonic and anharmonic chains (FPU chain and high
density LJ gas) have an eventual diffusive regime, after which the
effect of boundary sets (at times $t \approx L/c$) in, and size dependent oscillations appear.
These oscillations may be attributed to sound waves travelling in
the system. While they persist for a long time for harmonic chains,
they die off quickly for anharmonic chains.  At lower densities the
LJ gas behaves like the hard particle gas. We confirm this by
studying both equal and alternate mass LJ gas and see similar
behaviour as in \cite{roy13} for the hard particle gas.  Note that the transition of LJ gas from anharmonic chain like behaviour to hard-particle-like behaviour
is expected to be continuous and this would be interesting to study.
For the FPU chain, we find that a effective damped harmonic description of the sound modes reproduces the main observed features of the correlations quite well. 
We also revisited the simulations for the alternate mass hard particle gas studied in \cite{roy13} and presented results for larger system sizes.  We find that this exhibits an eventual diffusive behaviour.

In summary, we have studied tagged particle diffusion in various
one-dimensional Hamiltonian systems  and have come up with following results:
\begin{enumerate}
\item There is a
``short time'' regime during which the tagged particle at the centre
does not feel the effect of the boundary and during this time, the
correlation functions have the same behaviour as for an infinite
system. The tagged particle motion is {\emph{always}} diffusive. The diffusion constant is, in many cases {\emph{but not all}}, given accurately by the formula $D=k_BT/(2 \rho c)$. This formula is exact for a harmonic chain and appears to be accurate for the symmetric FPU chain and also the alternate mass hard-particle gas. It is less accurate for the asymmetric FPU chain and a Lennard-Jones gas. In the ``short time'' regime  we also find:
\begin{enumerate}
\item For the harmonic chain, the  VAF $\sim \cos (\omega
t)/t^{1/2}$.
\item For the FPU chain, the VAF has a faster decay $\sim \exp (-a t) \sin (\omega t)$.
\item The equal mass LJ gas, at high density, behaves like the FPU chain. At
low density it behaves like the equal mass hard-particle gas with
a negative VAF  $\sim -1/t^3$.
\item The alternate mass LJ gas at low density is also similar
to the alternate mass hard-particle gas with the VAF apparently changing
from $ -1/t^{3}$ to $ -1/t^{7/5}$.
\item The alternate mass hard particle gas also
 shows normal diffusive behaviour, but to see this asymptotic behaviour requires one to study very large system sizes. The VAF $\sim -1/t^{7/5}$. 
\end{enumerate}
\item At long times the effect of boundary sets in and size dependent
oscillations, damped for anharmonic chains, appear. The time at
which the system size effects start showing up is  $ \approx L/c$, where $c$ is the velocity of sound in the medium.

\end{enumerate}

An understanding of the rather striking observed differences in tagged particle correlations, between the FPU system and the alternating mass hard particle gas, remains an interesting open problem.

\end{document}